\documentclass[aps,pre,reprint]{revtex4-1}

\usepackage{graphicx}
\usepackage{amssymb,latexsym,amsthm,amsmath,bm}
\usepackage{natbib}
\begin{document}

\title{Relay synchronization in multiplex networks}

\author{I. Leyva}
\affiliation{Complex Systems Group {\& GISC}, Universidad  Rey Juan Carlos, 28933 M\'ostoles, Madrid, Spain}
\affiliation{Center for Biomedical Technology, Universidad Polit\'ecnica de Madrid, 28223 Pozuelo de Alarc\'on, Madrid, Spain}

\author{I. Sendi\~na-Nadal}
\affiliation{Complex Systems Group {\& GISC}, Universidad  Rey Juan Carlos, 28933 M\'ostoles, Madrid, Spain}
\affiliation{Center for Biomedical Technology, Universidad Polit\'ecnica de Madrid, 28223 Pozuelo de Alarc\'on, Madrid, Spain}

\author{R. Sevilla-Escoboza}
\affiliation{Centro Universitario de los Lagos, Universidad de Guadalajara, Jalisco 47460, Mexico}

\author{V.P. Vera-Avila}
\affiliation{Centro Universitario de los Lagos, Universidad de Guadalajara, Jalisco 47460, Mexico}

\author{P. Chholak}
\affiliation{Center for Biomedical Technology, Universidad Polit\'ecnica de Madrid, 28223 Pozuelo de Alarc\'on, Madrid, Spain}
\affiliation{Department of Mechanical Engineering, Indian Institute of Technology Bombay, Powai, Mumbai 400076, India}

\author{S. Boccaletti}
\affiliation{CNR-Institute of complex systems, Via Madonna del Piano 10, 50019 Sesto Fiorentino, Italy}
\affiliation{The Italian Embassy in Israel, Hamered Street 25, 68125 Tel Aviv, Israel}

\begin{abstract}
Relay (or remote) synchronization between two not directly connected oscillators in a network is an important feature allowing distant coordination. In this work, we report a systematic study of this phenomenon in multiplex networks, where inter-layer synchronization occurs between distant layers mediated by a relay layer that acts as a transmitter.
We show that this transmission can be extended to higher order relay configurations, provided symmetry conditions are preserved. By first order perturbative analysis, we identify the dynamical and topological dependencies of relay synchronization in a multiplex. We find that the relay synchronization threshold is considerably reduced in a multiplex configuration, and that such synchronous state is mostly supported by the lower degree nodes of the outer layers, while hubs can be de-multiplexed without affecting overall coherence. Finally, we experimentally validated the analytical and numerical findings by means of a multiplex of three layers of electronic circuits.
\end{abstract}
\maketitle
\thispagestyle{empty}

\section{Introduction}
Synchronization is one of the most important collective phenomena in nature. It can be observed in natural, social and technological systems, and  it became one of the most active research topics in network science \cite{Boccaletti2006,Arenas2008}. The huge amount of new data collected in the last years has permitted a higher resolution network representation of real systems. In particular, the inclusion of new features shaped multi-layer representations, i.e. approaches in which the network units are arranged in several layers, each one accounting for a different kind of interactions among the nodes \cite{DeDomenico2013,Boccaletti2014}. Multi-layer structures determine scenarios where novel forms of synchronization are relevant. Despite an analytical approach has been tackled in just a few particular cases  \cite{Sorrentino2012,Aguirre2014}, several synchronization scenarios have been already addressed, as unidirectional coordination between layers \cite{Gutierrez2012}, explosive synchronization emerging from the interactions between dynamical processes in multiplex networks \cite{Zhang2015,Nicosia2017},
complete synchronization \cite{delGenio2016,DSouza2016}, cluster synchronization \cite{Louzada2013,Jalan2016,Jalan2017}, intra-layer  \cite{Gambuzza2015} or inter-layer \cite{Sevilla2016,Leyva2017} synchronization.

Very recently, relay (RS) and remote synchronization (two very well known phenomena in chains, or small motifs, of coupled oscillators) have captured the attention of researchers. This form of synchronization is observed when two units of a network (identical or slightly different) synchronize despite not being directly linked, and due instead to the intermediation of a relay mismatched unit. The phenomenon has been experimentally detected in lasers \cite{Fischer2006} and circuits \cite{Bergner2012,Banerjee2012}. In general, the relay units exhibit generalized or delay synchronization with the units they actually pace to synchrony \cite{Gutierrez2013}.

RS is of outstanding relevance in the brain: the thalamus acts as a relay between distant cortical areas through the thalamo-cortical pathways, playing the role of a coordination hub that maintains the information flow \cite{Guillery2002,Sherman2007,Mitchell2014,Vlasov2017}. Complex structures and neuronal dynamics are implicated in this process involving not only simple, but higher order relay paths, that transfer the information through multiple-step relay chains \cite{Sherman2007,Mitchell2014}. Recently, remote synchronization has been addressed in the context of complex networks \cite{Gambuzza2016}, revealing the extremely important role of network structural and dynamical symmetries in the appearance of distant synchronization \cite{Nicosia2013,Pecora2014,Zhang2017}, as it was already suggested by the observation of zero-lag delays between mirror areas of the brain \cite{Konigqt1997,Soteropoulos2006}. Nevertheless, the interplay between symmetry, dynamics and multi-layer structure remains still mostly unexplored.

In this work, we perform a systematic study of inter-layer relay synchronization in a multiplex network, where distant layers synchronize their dynamics while their intra-layer motion remains unsynchronized. We consider generic high-order structures where multi-site relay pathways are verified. The dynamical and topological dependencies of the phenomena are studied, using perturbation stability analysis. The robustness of the relay synchronization against de-multiplexing the layers is reported, revealing the key role of low degree nodes in maintaining the layers coordination. Finally, the findings are experimentally validated in a multiplex network of electronic circuits.

\section{Results}

\subsection{Model}

\begin{figure}[t]
 \centering 
   \includegraphics[width=0.4\textwidth]{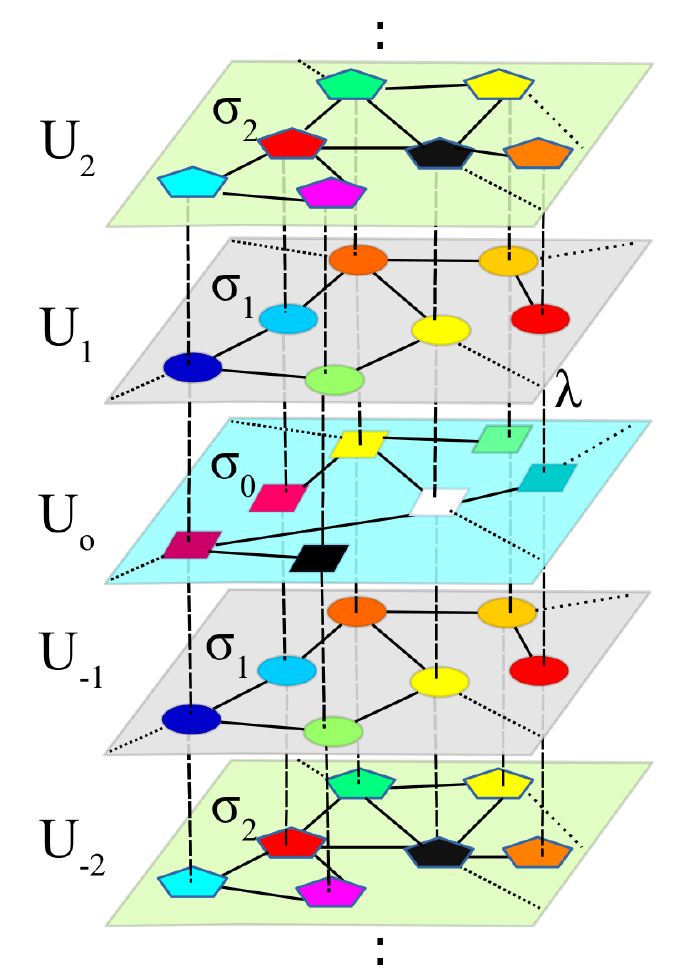}
\caption[]{ Schematic representation of a multiplex of $2M+1$ layers (here $M=2$) labeled as $k=-M,\dots,-1,0,1,\dots,M$ where each pair of layers $k$ and $-k$ (painted with the same color) are networks of identical oscillators with the same topology ${\cal L}^k$ and intra-layer coupling $\sigma_k$ and whose dynamical state is described by the variable ${\bf U}^k$ and ${\bf U}^{-k}$, respectively. The multiplex is symmetric with respect to the layer $k=0$ and  the nodes are coupled to their replicas in the rest of layers with an inter-layer coupling strength $\lambda$.  \label{fig1}}
\end{figure}

We start by considering $2M+1$ layers (or networks), arranged as shown in Fig.~\ref{fig1}. Each layer $k$, with $k=-M,\dots,0,\dots,M$, is formed by $N$ $m$-dimensional dynamical systems whose states are represented by the column vectors ${\bf U}^{k}=\{{\bf u}_{1}^{k},{\bf u}_{2}^{k},\ldots,{\bf u}_{N}^{k}\}$, and whose intra-layer interactions are encoded through the Laplacian matrices  $\mathcal{L}^k=\lbrace\mathcal{L}^{k}_{ij}\rbrace$. The layer stack is symmetric with respect to $k=0$ in such a way that Laplacians $\mathcal{L}^k$ and $\mathcal{L}^{-k}$ have the same structure. The dynamical systems are also paired: nodes belonging to layers ${\bf U}^{+k}$ and ${\bf U}^{-k}$ are identical to each other, and different (in some parameter) from the rest of the layers. Consequently, layer $k=0$ has no counterpart, and acts as a relay between all layers situated above and below it.

 Layers are coupled in a multiplex configuration, and the dynamical evolution of the system is described by the following equations:
\begin{footnotesize}
\begin{equation}
  \label{eq:multiplex}
\dot {\bf U}^k = {\bf F}_k({\bf U}^k) - \sigma_k(\mathcal{L}^k \otimes  {\bf G}) {\bf U}^k +  \lambda ({\bf \mathbb{I}_N \otimes H}) \sum_{q=k-1\geq -M}^{q=k+1\leq M} \left({\bf U}^q-{\bf U}^k\right)
\end{equation}
\end{footnotesize}

\noindent where the functions ${\bf F}_k({\bf U}^k)=[{\bf f_k}({\bf u}_1^{k}),{\bf f_k}({\bf u}_2^{k}),\ldots,{\bf f_k}({\bf u}_N^{k})]^T$ (with ${\bf f_k}:\mathbb{R}^{m}\!\to\!\mathbb{R}^{m}$ representing the vectorial functions evolving each dynamical unit),  are identical for the same $|k|$.  ${\bf G},{\bf H}$ are the $m \times m$ matrices representing respectively the linear intra- (${\bf G}$)  and inter- (${\bf H}$) layer coupling schemes.  ${\bf \mathbb{I}_N}$ is the $N \times N $ identity matrix, $\sigma_k$ is the intra-layer coupling strength within  layers $k$ and $-k$,  and $\lambda$ is the inter-layer coupling strength.

Due to the reflection symmetry of the system under study (i.e. as long as the ${\bf U}^{+k}$  and ${\bf U}^{-k}$ layers are identical for all $k$), a synchronous inter-layer evolution (with layers evolving in a pairwise synchronized fashion, i.e. where ${\bf U}^{+k}={\bf U}^{-k}$) at all $k$ without necessarily implying ${\bf U}^{k} = {\bf U}^{k'}$ for $k \neq k'$) is a solution of Eqs.~(\ref{eq:multiplex}), independently of intra-layer synchronization \cite{Sevilla2016} (i.e. independently on whether the state of the systems within each layers are synchronized).
Let therefore $ {{\delta {\bf U}^k}}(t) = {\bf U}^{+k}(t) - {\bf U}^{-k}(t)$, with $k=1,\dots,M$ be the vector describing the difference between the dynamics of the paired layers.

 Considering the smallness of ${\delta{\bf U}^k}=\{{ \delta {\bf u}^k_1},{\delta {\bf u}^k_2},\ldots,{\delta {\bf u}^k_N}\}$  and expanding around the inter-layer solution up to first order, one  obtains a set of $M$ linearized vector equations for the perturbations $\delta {\bf U}^k$:
\begin{align}\nonumber
\label{eq:variational_id}
 \dot{\delta {\bf U}}^k & =  \left[ J{\bf f}(\tilde{\bf U}^k) -\sigma_k (\mathcal{L}^k \otimes J{\bf G}) - \lambda(2-\delta_{kM}) J {\bf H}(\tilde{\bf U}^k)\right] \delta {\bf U}^k  \\
& +\lambda \sum_{\substack{q=k-1 \\q\neq k}}^{q=k+1 } J{\bf H}(\tilde{\bf U}^q) \delta{\bf U}^q
\end{align}

\noindent where $J$ denotes the Jacobian operator, $\delta_{kM}$ is the Kronecker delta accounting for the boundary condition at $k=M$ (as the stack end layers ${\bm U}^{\pm M}$ are only connected to the previous neighbor layer). The vector $\tilde  {\bm U}^k=\left\lbrace {\tilde {\bm u}^{k}_{i}} \right\rbrace$ describes the dynamical state of any of the $k=0,\dots,M$ layers at the synchronous state ${\bf U}^k={\bf U}^{-k}\neq {\bf U}^0$ and, therefore, the whole dynamics is reduced to the dynamics of the  $M+1$ layers.

 Such evolution at the node level is given by:
\begin{equation}
\label{eq:solution_nodes}
{\bf\dot{\tilde{u}}}^{k}_{i} = {\bf f_k}({\bf\tilde{u}}^{k}_{i}) - \sigma_k \sum_j\mathcal{L}^{k}_{ij}\,{\bf g}({\bf\tilde{\bf u}}^{k}_{j}) +\lambda \sum_{q=k-1\geq 0}^{q=k+1\leq M} \left[]{\bf h}({\bf\tilde{u}}^{q}_{i})- {\bf h}({\bf\tilde{u}}^{k}_{i})\right]
\end{equation}
 where $i=1\dots,N$ is the node index, and $k,q=0,\dots,M$. Notice that, since each paired layers $k$ and $-k$ is inter-layer synchronized ($\tilde{\bf U}^k={\bf U}^{k}={\bf U}^{-k}$), each layer acts therefore as a relay to the rest of the stack. The $Mm$ linear equations (\ref{eq:variational_id}), solved in parallel to the $(M+1)m$ nonlinear equations (\ref{eq:solution_nodes}) for ${\bf\dot{\tilde u}}^k_i$, allow for calculating all Lyapunov exponents transverse to the manifold $\tilde{\bf U}^k$.  The maximum of those exponents (MLE) as a function of the system parameters actually gives  the necessary conditions for the stability of the synchronous solution: whenever $\text{MLE}<0$, perturbations transverse to the manifold will die out, and the multi-relay synchronous solution will be stable.

\begin{figure}[t]
  \centering
  \includegraphics[width=0.45\textwidth]{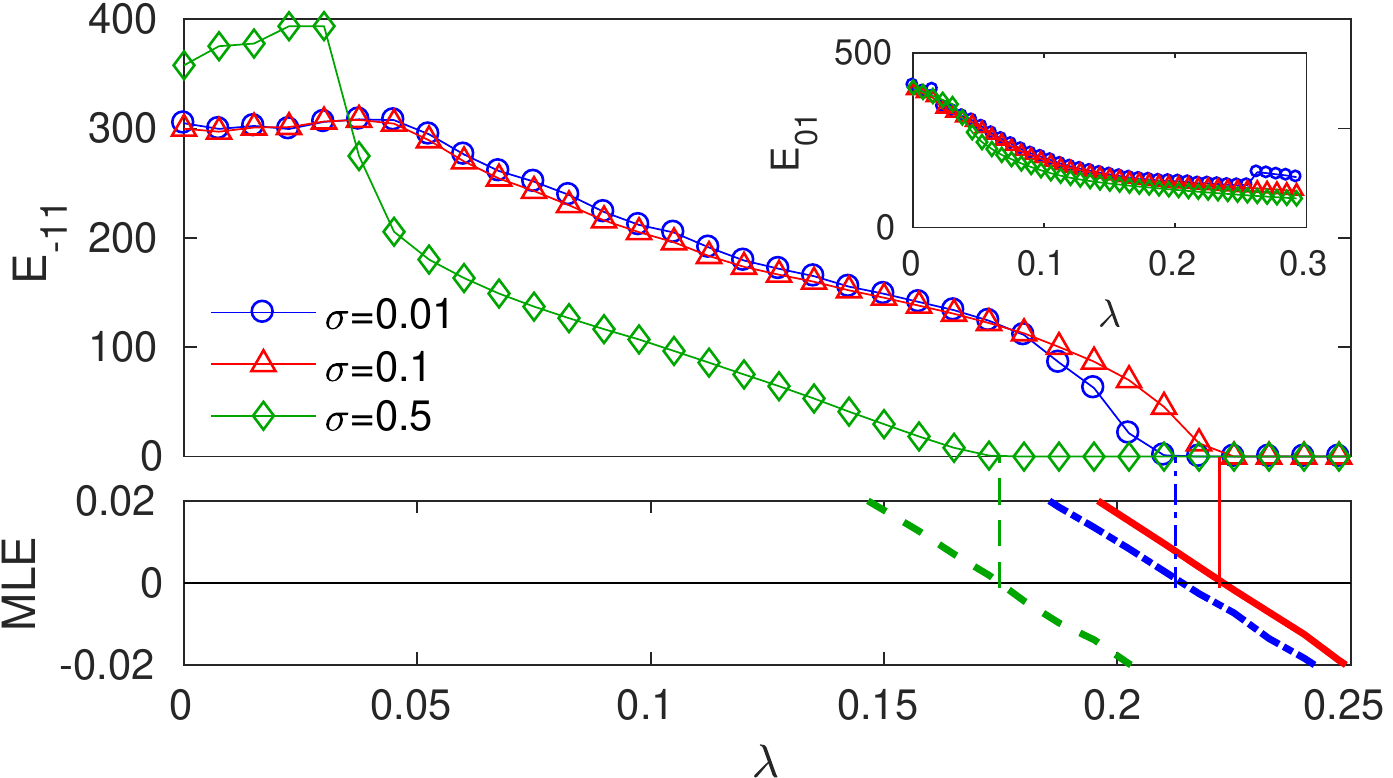}
\caption[]{Relay synchronization in a triplex ($M=1$) with identical SF layers (SSS configuration). (Main panel) Synchronization error between the outer layers ($k=-1$, and $k=1$) $E_{-11}$ (see Eq. \ref{eq:Eintra}) as a function of the inter-layer coupling $\lambda$ for three different values of the intra-layer coupling $\sigma_0=\sigma_1$ (see legend). The inset shows the corresponding synchronization errors between the relay and one of the outer layers. (Bottom panel) Maximum Lyapunov exponent (MLE) of the relay synchronization manifold $U^1=U^{-1}$ as a function of $\lambda$ for the same cases as in the main panel. Vertical lines mark the points where the MLE becomes negative. All points are averages of $10$ network realizations with $N=500$ and $\langle k\rangle=4$. See the main text for the relay and outer layer R\"ossler oscillators specifications.}
\label{fig2}
\end{figure}

In order to monitor the synchronization error between layers, we define the inter-layer synchronization error as,
 \begin{equation}
   \label{eq:Eintra}
 E_{qk}=\lim_{T\to \infty}\frac{1}{T}\int_0^T \sum_{i=1}^{N} \left\lVert {\bf u}^{q}_{i}(t)-{\bf u}^{k}_{i}(t)\right\rVert dt,
  \end{equation}
where $\lVert \cdot \rVert$ stands for the Euclidean norm and $q,k$ are the layers' indexes, such that $E_{-kk}$ denotes the inter-layer synchronization error of mirror layers. Without lack of generality, in our numerical simulations we consider two types of topologies where layers are either (i) Erd\"os-Renyi \cite{erdos1959} (ER) or (ii) scale-free \cite{Barabasi1999} (SF), in all cases with $N=500$. We classify  the layer stacks regarding the topology sequence of each layer. For instance, a triplex where the three layers have ER topology will be denoted as EEE, and a system where two identical SF layers are mediated by a center ER will be denoted as SES. The nodes are chaotic R\"ossler oscillators \cite{Rossler1976}, defined by  the $m=3$ state vector ${\bf u}=(x,y,z)$ whose autonomous evolution is given by ${\bf f_k}({\bf u})={\bf f_{-k}}({\bf u})=\left[-y-z,x+a_k y, 0.2+z(x-9)\right]$ and the heterogeneity between the layers is introduced through the parameter $a_k$.
In our  case study, the intra- and inter- layer coupling functions are set to be $\mathbf{g}({\bf u})={\bf G u}=(0,0,z)^T$ and  $\mathbf{{h}}({\bf u})={\bf Hu}=(0,y,0)^T$  respectively. These coupling schemes ensure that intra-layer synchronization is prevented when layers are isolated and not multiplexed (class I layers, according to the standard master stability function (MSF) classification established in Ref.~\cite{Boccaletti2006}) whereas multiplexed nodes along the layers can synchronize for a coupling strength $\lambda$ above a given threshold (class II MSF).

\subsection{Layers with identical topology}
\label{sub:identical}

With the aim of determining whether relay synchronization can be achieved in a multiplex configuration let us first consider the multiplex structure defined by Eq.~(\ref{eq:multiplex}) for the case of three identical SF layers ${\cal L}^0={\cal L}^1={\cal L }^{-1}$ and where the parameters $a_1=a_{-1}=0.2$ for the outer layers and $a_0=0.3$ for the relay units of the central layer, although different selections of these parameters and topologies produce a similar behavior.

Results are collected in Fig.~\ref{fig2}, where the synchronization error between the outer layers $E_{-11}$ is plotted versus the inter-layer coupling $\lambda$ for several values of the intra-layer couplings $\sigma_1$ and $\sigma_0$ in the outer and relay layers respectively, with $\sigma_1=\sigma_0$. In all cases, there is a critical coupling $\lambda^*$ above which complete synchronization between layers $k=1$ and $k=-1$ occurs, that is, $E_{-11}=0$ is achieved for any generic initial condition and network realization, while  the relay layer ($k=0$) remains unsynchronized to any of the two outer layers ($k=1,-1$) as shown in the inset where $E_{01}>0$ for any parameter choice.

\begin{figure}[t]
  \centering
   \includegraphics[width=0.5\textwidth]{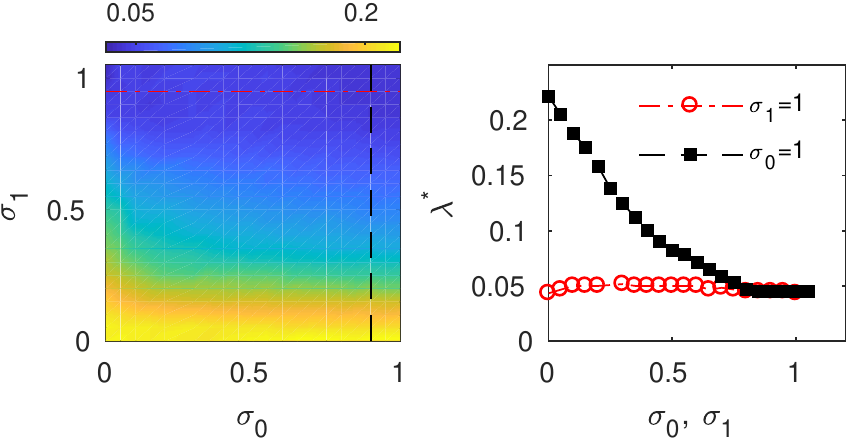}
\caption[]{Relay synchronization in a triplex network with identical SF layers as a function of the intra-layer couplings for the relay ($\sigma_0$) and outer ($\sigma_1$) layers. (Left) Color map of the inter-layer coupling threshold $\lambda^*$ for the relay state ($E_{-11}=0$ and $E_{01}\neq 0$) in the $\sigma_0$-$\sigma_1$ parameter space. (Right) Inter-layer coupling threshold $\lambda^*$ for the relay state as a function of the coupling strength in the relay layer $\sigma_0$ for a fixed value of $\sigma_1=1.$ (red dashed line in left panel) and as a function of the coupling strength in the outer layers $\sigma_1$ for a fixed value of $\sigma_0=1.$ (black dashed line in left panel).    Each point is an average of 10 SF network realizations with $N=500$ and $\langle k\rangle=8$. \label{fig3}}
\end{figure}
In addition, the calculation of the corresponding MLE given by Eqs.~(\ref{eq:variational_id}) (lower panel of Fig.~\ref{fig2}) confirms that the relay synchronous solution ${\bf U}^{-1}={\bf U}^1$ reaches stability ($\text{MLE}<0$) at the same critical $\lambda^*$ where the error between the relay and the outer layers is zero, as indicated by the vertical dashed lines. Therefore, one can conclude that inter-layer MLE is a useful tool for reducing the system's dimensionality and use it for evaluation of the critical inter-layer coupling  $\lambda^*$ from now on.

In order to better understand the different roles played by external and relay layers, we show in Fig.~\ref{fig3} the critical inter-layer coupling value in the parameter region $(\sigma_0, \sigma_1)$, that is, when the intra-layer coupling $\sigma_k$ is different for the relay and outer layers. It can be seen that the system's ability to synchronize is practically unaltered with $\sigma_0$, while increasing $\sigma_1$ makes the value of $\lambda^*$ to drop drastically. This therefore reveals that multiplex relay synchronization is much more sensitive to changes affecting the mirror layers than to those arising in the transmission layer.

\begin{figure}
 \centering
  \includegraphics[width=0.45\textwidth]{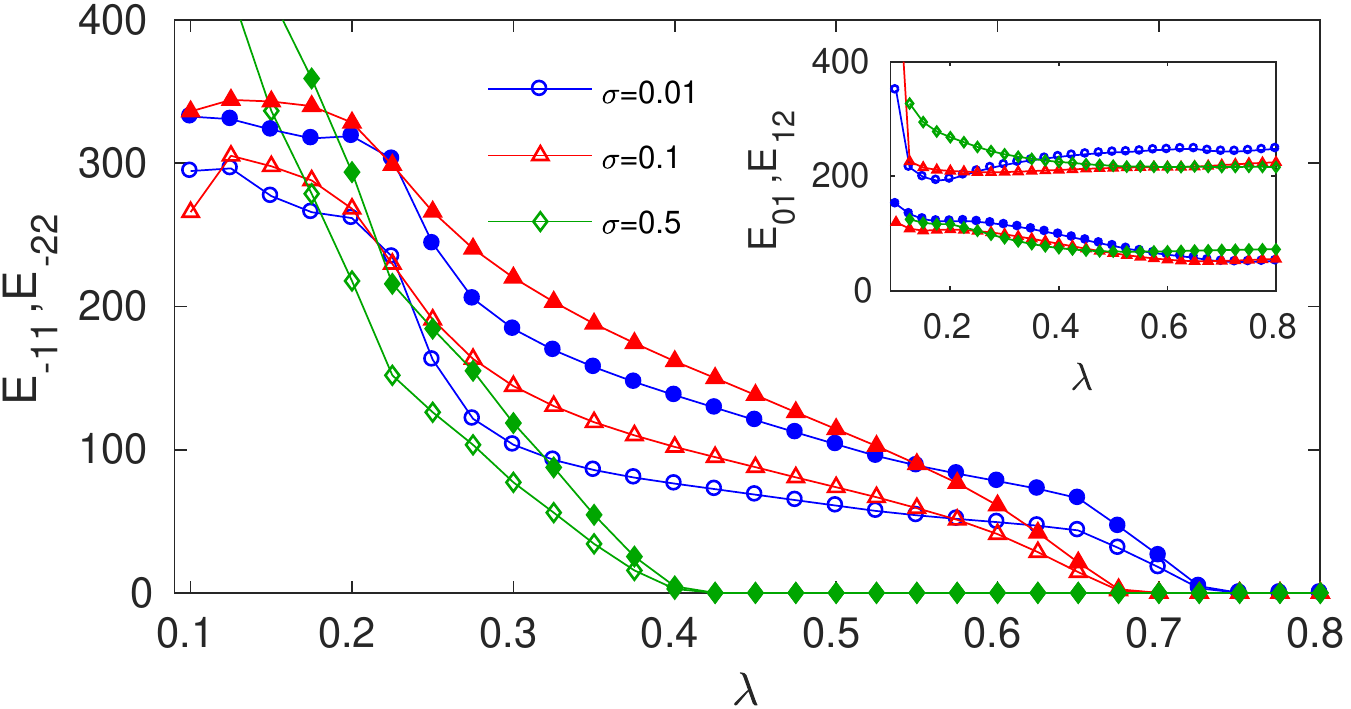}
\caption[]{Relay synchronization in a pentaplex ($M=2$) with identical $N=500$ ER layers (EEEEE configuration). The synchronization error between the two pair of outer layers $E_{-11}$ (empty symbols) and $E_{-22}$ (full symbols) is shown as a function of $\lambda$ for three different values of the intra-layer coupling $\sigma$, being $\sigma=\sigma_k,$ $\forall k$. The inset shows the synchronization errors between each one of the outer layers and the relay layer.
}\label{figpenta}
\end{figure}

Our results can be generalized to any number of layers. As an example, we report also the case $M=2$, which corresponds to two outer layers above ($k=1,2$) and below ($k=-1,-2$) the relay layer ($k=0$). We choose $a_{-1}=a_{1}=0.2$ and  $a_{-2}=a_{2}=0.3$, and $a_0=0.25$ for the central layer. The results stand for any other parameter choice. In Fig.~\ref{figpenta} we plot the inter-layer synchronization errors $E_{-11}$ (void markers) and  $E_{-22}$ (full markers), vs. the inter-layer coupling $\lambda$ for several values of the intra-layer coupling $\sigma$. As in the triplex case, the critical $\lambda^*$  at which complete inter-layer synchronization is achieved depends on $\sigma$, but it is the same for both pairs of layers, as  $E_{-11}$ and  $E_{-22}$ drop to zero simultaneously. In the inset we plot the inter-layer synchronization errors between the non-paired layers, $E_{01}$, $E_{12}$ to check that they remain mutually incoherent. Therefore, we have verified that relay synchronization also occurs in cascade for arbitrarily high-order multiplex systems, provided a structural and dynamical symmetry is conserved.

\subsection{Layers with non-identical topology}
\label{sub:nonidentical}

\begin{figure}
  \centering
  \includegraphics[width=0.5\textwidth]{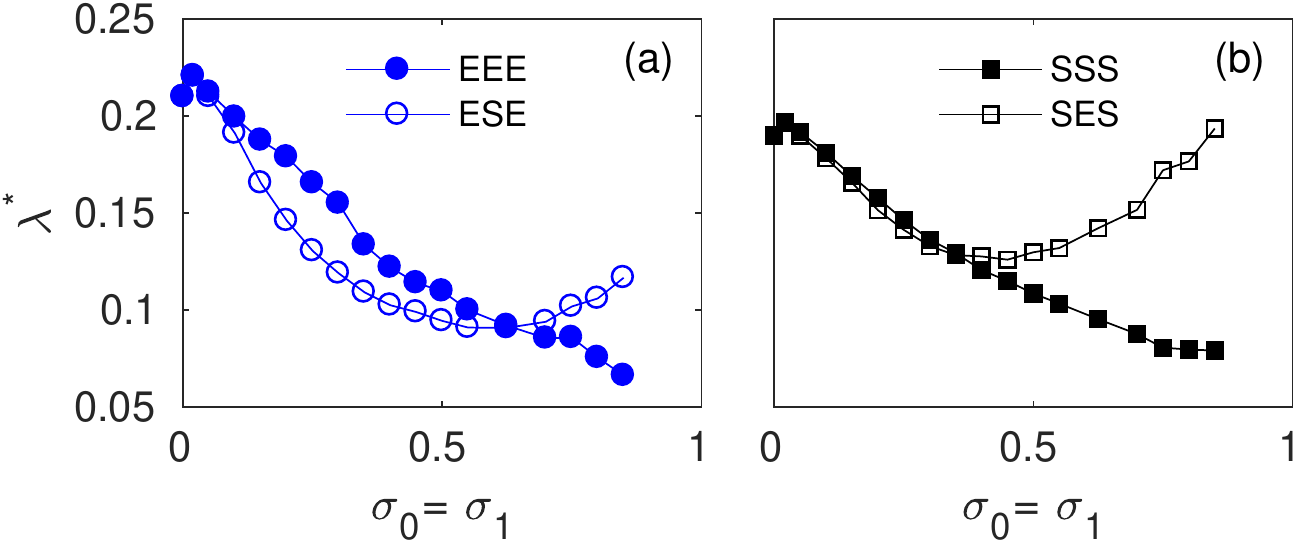}
\caption[]{Relay synchronization in a triplex with different layers. Inter-layer coupling threshold $\lambda^*$ vs the intra-layer couplings $\sigma_0=\sigma_1$ for (a) a mixed ER-SF-ER (ESE) and identical (EEE) configurations and (b) a mixed SF-ER-SF (SES) and identical (SSS) configurations.
\label{fig4}
}
\end{figure}

So far, we have dealt with multiplexes of pairwise identical layers. However, this condition is too strong a limitation to hope that it would capture and properly represent the case of many real systems. The next step needed for generalization is studying then the relay synchronization scenario in the case in which the topology of the relay layer differs from that of the outer layers. In Fig.~\ref{fig4} we  have reported  the critical inter-layer coupling  $\lambda^*$ in two heterogeneous triplex cases: (a) a pair of Erd\"os-R\'enyi  layers mediated by a scale-free relay layer (ESE situation) and (b) the opposite case where SF layers are connected through a ER layer (SES).  Each case is compared with the topologically homogeneous  EEE and SSS structures, respectively. For the sake of simplification and of better assessment of the role of the topology, we keep $\sigma_0=\sigma_1$.

Figure \ref{fig4}(a) shows that, for a large range of intra-layer couplings, the mediation of a SF relay facilitates the synchronization between the paired layers, since $\lambda^*$ in the ESE case (void blue circles) is smaller than the one corresponding to the homogeneous case (EEE, full blue circles). On the contrary, a relay ER layer intermediating between two outer SF layers  (Fig.~\ref{fig4}(b)) does not determine a significant difference as long as the intra-layer coupling strength is low, but increases the threshold $\lambda^*$ for higher $\sigma$, as compared to the homogeneous SSS case.

\subsection{Robustness}
\begin{figure}[t]
  \centering
  \includegraphics[width=0.23\textwidth]{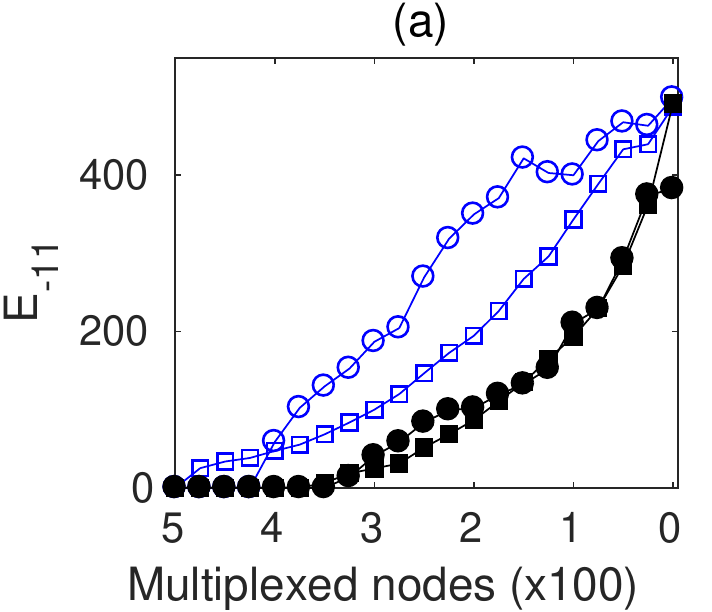}
  \includegraphics[width=0.23\textwidth]{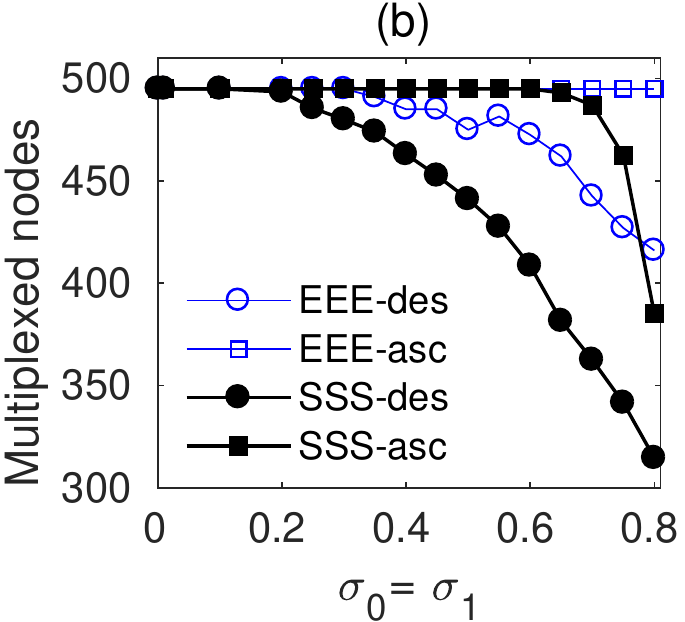}
\caption[]{Robustness of the network relay synchronization for identical layers. (a) Synchronization error between the outer layers $E_{-11}$ vs the decreasing number of connected relay lines for identical ER (blue empty symbols) and SF (black solid symbols) layers. Relay lines are disconnected following a descending (circle symbols) or ascending (square symbols) node degree ranking of the outer layers (seed legend in (b)). Parameter values are $N=500$, $\langle k\rangle 8$, $\lambda=0.23$ and $\sigma_0=\sigma_1=0.8$. (b) Number of multiplexed relay lines needed to support a relay network state as a function of the intra-layer coupling strength $\sigma_0=\sigma_1$ while keeping constant $\lambda=0.23$. The different curves are explained in the legend.}
\label{fig5}
\end{figure}
In the previous Sections we have addressed the dependence of relay synchronization in a multiplex on the dynamical and structural layer heterogeneity, and proved that the phenomenon still holds even when the intermediate layer has a completely different structure and dynamics than the mirrored ones. The present section is devoted instead to assess the robustness of relay synchronization against a de-multiplexing process of the layers, that is, against performing a progressively shutting down of the inter-layer links such that a fraction of nodes in each layer is not linked to their counterparts in the other layers.

To closely check this process, we initially consider a 3-layer multiplex with identical topology (EEE or SSS). We choose the inter- and intra-layer couplings to guarantee a relay synchronous state with the layers fully multiplexed. Then, we proceed to disconnect one by one the inter-layer links according to the nodes degree ranking, both in the ascending and the descending order, and re-evaluate in every step the state of the relay synchronization by measuring the $E_{-11}$ error. An example of such a process is shown in Fig.~\ref{fig5}(a) by reporting the evolution of $E_{-11}$ as a function of the number of multiplexed nodes. It can be seen that, starting from a situation with $E_{-11}=0$, the EEE multiplex configuration (blue void markers) looses the synchronization immediately with just a few of inter-layer links being removed. On the other hand, relay synchronization is resilient in SSS triplex configurations also when more than $30\%$ of the nodes are not multiplexed.

A more detailed view can be obtained from Fig.~\ref{fig5}(b), where the number of multiplexed nodes needed to support the relay synchronization is represented as a function of the intra-layer coupling $\sigma_0=\sigma_1$. As expected, when the coupling is weak, all the $N$ nodes need to be linked to preserve relay synchronization. However, as the interaction within the layers increases, the intra-layer connectivity helps to maintain a synchronous state despite an increasing number of nodes are being de-multiplexed without damaging the coherence between the outer layers. In Fig.~\ref{fig5}(b), we can see that for both the EEE (blue void markers) and the SSS (black full markers) triplex configurations, removing the links between layers connecting nodes with higher degree (descending degree ranking, circle markers) is much more robust than following an ascending degree ranking (square markers). This is indeed a very interesting result: relay synchronization in a multiplex network is supported by the low degree nodes, while the hubs can be safely disconnected without perturbing the transmission. This is notably evidenced in the SSS case (black full squares) where after having removed the $40\%$ of the inter-layer links connecting the highest degree nodes, the relay synchronization is still supported by the multiplex structure connected through the lowest degree nodes.
\begin{figure}[t]
  \centering
  \includegraphics[width=0.25\textwidth]{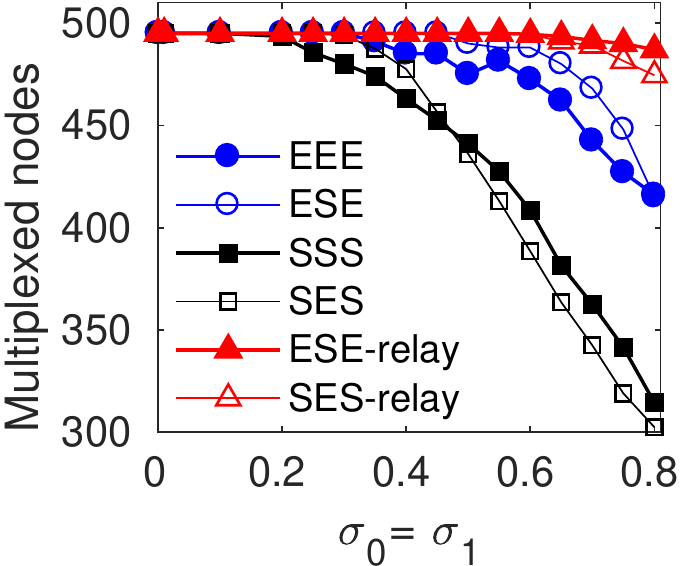}
\caption[]{Robustness of the network relay synchronization for non identical layers. Number of multiplexed relay lines needed to support a relay network state as a function of the intra-layer coupling strength $\sigma_0=\sigma_1$ while keeping constant the inter-layer coupling strength $\lambda=0.23$ for mixed ER (ESE, empty circles) and mixed SF (SES, empty squares) layer configurations. The relay lines are disconnected following a descending order of the outer node degrees and for comparison the corresponding values for identical layers -see Fig.~\ref{fig5}(b) are plotted in solid symbols. The red solid (ESE) and empty (SES) triangles show the behavior when the relay lines are disconnected following the degree ranking of the relay layer.}
\label{fig6}
\end{figure}

Once we have singled out the descending degree ranking as the most convenient way to de-multiplex part of the network without loosing coherence, we proceed our study by evaluating the impact of having a relay layer with different topology from the outer layers, as we did in the previous Section~\ref{sub:nonidentical}. In this scenario, we have two possible descending degree rankings, the one dictated by the relay layer and the one dictated by the outer layers. The results are summarized in Fig.~\ref{fig6} where we plot, as in Fig.~\ref{fig5}(b), the  number of nodes that need to be linked to maintain synchronization as a function of $\sigma_0=\sigma_1$. For the sake of comparison, we added the curves for the homogeneous EEE and SSS (full markers) multiplex configurations, together with the data for the mixed ESE and SES (void markers) layers. Notice that the chosen inter-layer coupling $\lambda=0.23$ is well above threshold for all the cases, as it can be derived from Fig.~\ref{fig4}.
All the reported evidence indicates that the introduction of a relay layer with a topology different from that of the outer layers has little influence on the minimum number needed to support the relay synchronization, as long as the first removed inter-layer connections correspond to the hubs in the outer layers (blue and back curves). Curiously, the alternative of using the relay layer topology to rank the degree of the nodes, destroys the coherence between the outer layers as soon as a tiny fraction of links is removed (red curves). Therefore, the relay synchronization in a multiplex is very unstable if just a few links connecting nodes which are hubs in the relay layer are removed. Notice that this unlinking criterion is equivalent to randomly disconnect the multiplex. Therefore, the robustness of the relay synchrony relies mainly in the low degree nodes of the external layers. The relevance of the low degree nodes in controlling the dynamics of complex networks has been pointed out in other contexts \cite{Liu2011,Skardal2015}.

\section{Experimental validation}
\label{sub:experimental}

\begin{figure*}
  \centering
  \includegraphics[width=0.74\textwidth]{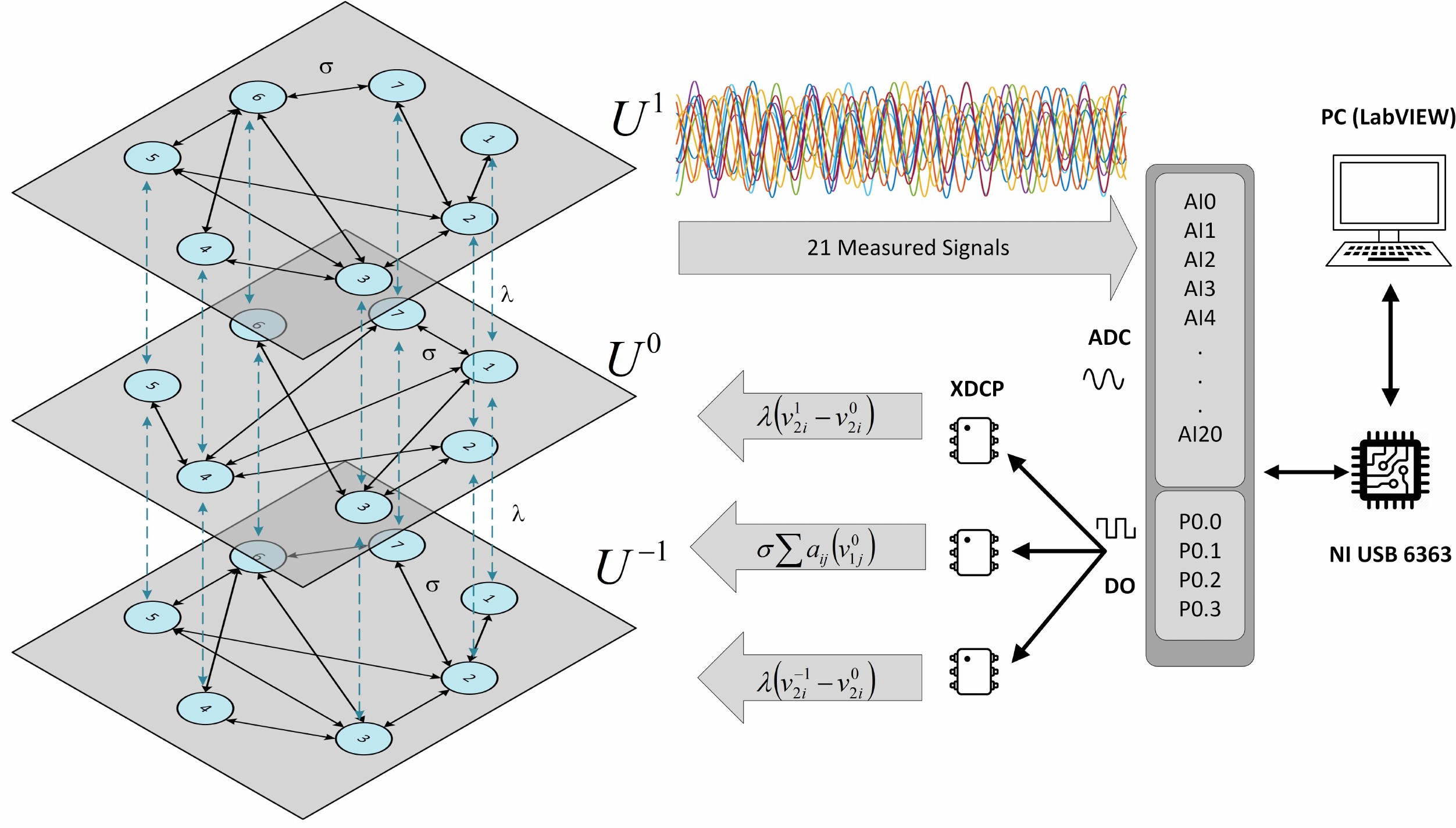}
  \includegraphics[width=0.25\textwidth]{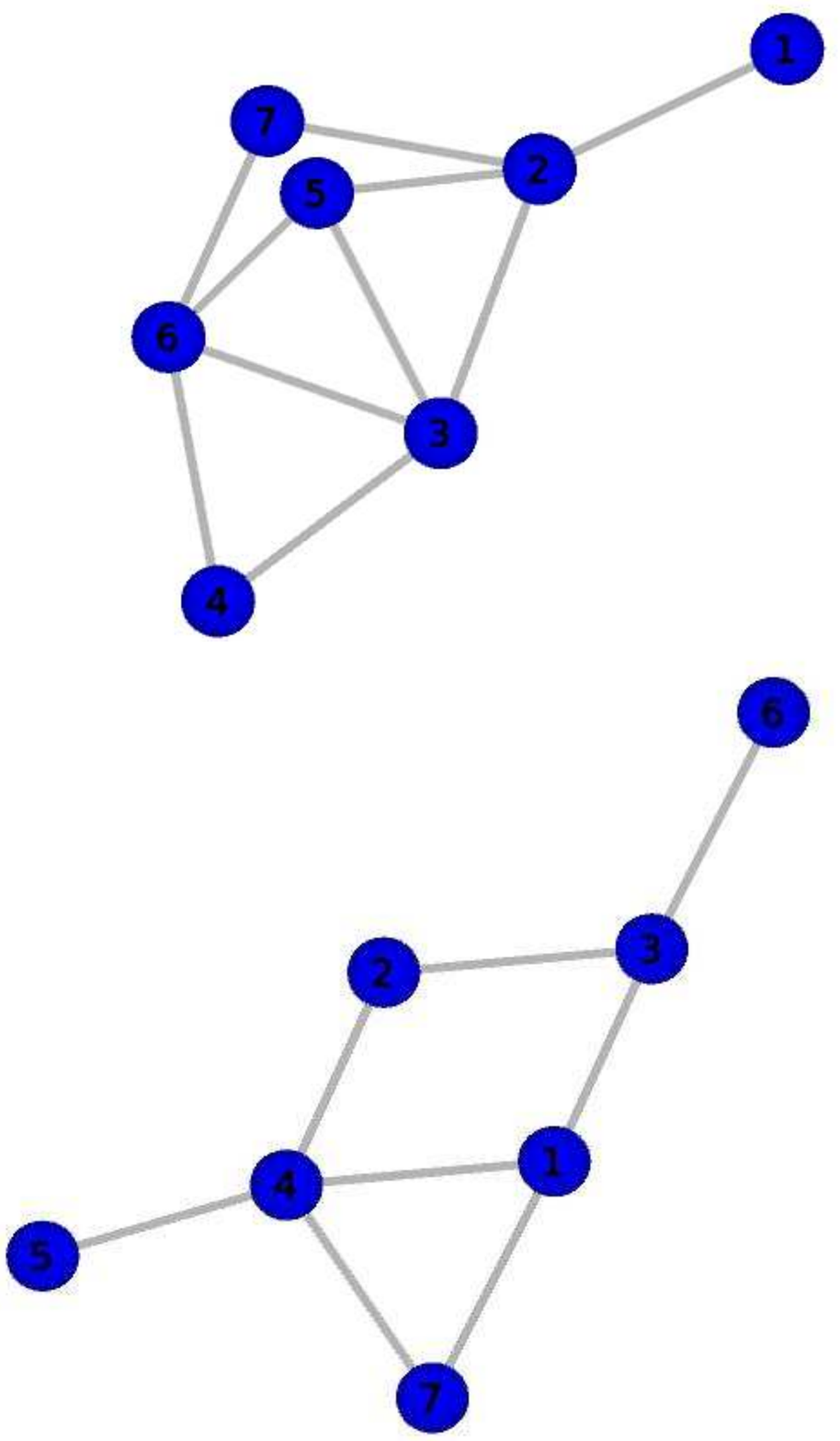}
\caption{(Left) Schematic representation of the experimental arrangement of three layers of electronic circuits. The bidirectional coupling is adjusted by means of three strips of digital potentiometers X9C103 (XDCP), the resistance is controlled through digital pulses sent by a DAQ (NI USB 6363). (Right) Graph structure used for the upper and lower layers (top) and for the relay layer (bottom).}
\label{fig:setup}
\end{figure*}

Finally, we report experimental evidence of relay synchronization in a multiplex of nonlinear electronic circuits, with the setup sketched in Fig.~\ref{fig:setup} (left).
The array is made of 21 R\"ossler-like circuits arranged in three layers of 7 nodes, with the relay layer having different topology as the outer layers. Each layer has two different electronic couplers, one for the coupling among nodes in the same layer ($\sigma_e$) and the second for the interaction of each node with its replica in the other layers ($\lambda_e$). The chaotic dynamics of the circuits is well approximated by the three variables $(v_{1},v_{2},v_{3})$ obeying \cite{Sevilla2016}:
\begin{eqnarray}
\label{eq:experimentalv}
\dot{v}_{1i}^{k} & = & -\frac{1}{R_{1}C_{1}}\left(v_{1i}^{k} + \frac{R_{1}}{R_{2}}v_{2i}^{k}+\frac{R_{1}}{R_{4}}v_{3i}^{k}\right)\\ \nonumber
&& -\frac{1}{R_{1}C_{1}}\sigma_{e}\frac{R_{1}}{R_{15}}\sum_{j=1}^{N}{a_{ij}^{k}(v_{1j}^{k}-v_{1i}^{k})}\\ \nonumber
\dot{v}_{2i}^{k} & = & -\frac{1}{R_{6}C_{2}}\left(-\frac{R_{6}R_{8}}{R_{9}R_{7}}v_{1i}^{k}+ \left[1- \frac{R_{6}R_{8}}{R_{c}^{k}R_{7}}\right]v_{2i}^{k}\right) \\ \nonumber
&& -\frac{1}{R_{6}C_{2}}\left(\lambda_{e}\frac{R_{6}}{R_{16}} \sum_{q=-1}^{q=1}{v_{2i}^{q}-v_{2i}^{k}}   \right)\\ \nonumber
\dot{v}_{3i}^{k} & = &-\frac{1}{R_{10}C_{3}}\left(-\frac{R_{10}}{R_{11}}G\left( v_{1i}^{k}\right)+v_{3i}^{k} \right)\\ \nonumber
\end{eqnarray}
where $G_{v_{1i}}$ is a nonlinear gain funtion given by:
\begin{eqnarray}
 G(v_{1i})&=& \begin{cases} 0, &                                                                                                  \mbox{if } v_{1} \le F(I) \\
            \frac{R_{12}}{R_{14}}v_{1i}-Vee\frac{R_{12}}{R_{13}}-Id\left(\frac{R_{12}}{R_{13}}+\frac{R_{12}}{R_{14}} \right), & \mbox{if } v_{1} > F(I) \end{cases} \nonumber \\
F(I) &=& Id(1+\frac{R_{14}}{R_{13}})+Vee\frac{R_{14}}{R_{13}}
\end{eqnarray}

where the parameter values are gathered in Table~\ref{paramvalues}. The reader is referred to Ref.~\cite{DBSevilla2017} for a detailed description of the experimental implementation of the R\"ossler-like circuit in the networks, and Refs.~\cite{Aguirre2014,Sevilla2015,Sevilla2016,Leyva2017} for previous realizations in different network configurations. Both the intra-layer $\sigma_e$ and the inter-layer $\lambda_e$ are set by means of the digital potenciometers X9C103, that working as voltage divisor for the maximum resitence (10k$\,\Omega$), $\sigma_e$ and $\lambda_e$ is set to zero, this potentiometers are controlled through the digital ports (P0.0, P0.1, P0.2, P0.3) of a DAQ card. First that all we send all the coupling value to zero, after 500ms takes the sample of the time series of each networks, all the variables $v_{2i}$ of each oscillator enter to the DAQ card through the analogue ports (AI0, AI1, $\dots$ , AI20) and saved in the PC for further analysis. Next, the coupling between the inter-layer ($\lambda_{e}$) increases  one step $(0.01)$, digital pulses are sent to the potenciometers corresponding to that coupling and decreases the resistance in $100\,\Omega$ each time it passes through this state, until the maximum value of $\lambda_{e}$ is reached ($\,\Omega$ in potenciometers). When  the entire run is finished, $\sigma_{e}$ is increased by one step, and another cycle of $\lambda$ is initiated. The whole procedure is repreted until each coupling reached its maximum value. The experiment is controlled from a PC with the LabVIEW software.
\begin{table}[b]
\caption{Parameter values of the chaotic dynamics of one R\"osller like circuit as described in
Eqs.~(\ref{eq:experimentalv}).\label{paramvalues}}
\begin{center}
\begin{tabular}{|l|l|l|l|}
\hline
	C1=1nF				& C2=1nF				& C3=1nF				& $\sigma_{e},\lambda_{e}=[0-0.6]$ \\
\hline
 $R_{1}=2\,M\Omega$ 	& $R_{2}=200\,k\Omega$& $R_{3}=10\,k\Omega$ 	& $R_{4}=100\,k\Omega$ \\
\hline
$R_{5}=50\,k\Omega$ & $R_{6}=\,5M\Omega$ 	& $R_{7}=100\,k\Omega$	& $R_{8}=10\,k\Omega$ \\
\hline
$R_{9}=10\,k\Omega$ & $R_{10}=100\,k\Omega$ & $R_{11}=100\,k\Omega$	& $R_{12}=150\,k\Omega$ \\
\hline
$R_{13}=68\,k\Omega$ &$R_{14}=10\,k\Omega$ & $R_{15}=75\,k\Omega$	&  $R_{16}=120\,k\Omega$  \\
\hline
$Rc^{0}=50\,k\Omega$& $Rc^{1}= 35\,k\Omega$ &$V_{d}=0.7$ 			& $V_{ee}=15$\\
\hline
\end{tabular}
\end{center}
\end{table}


\begin{figure}
  \centering
  \includegraphics[width=0.45\textwidth]{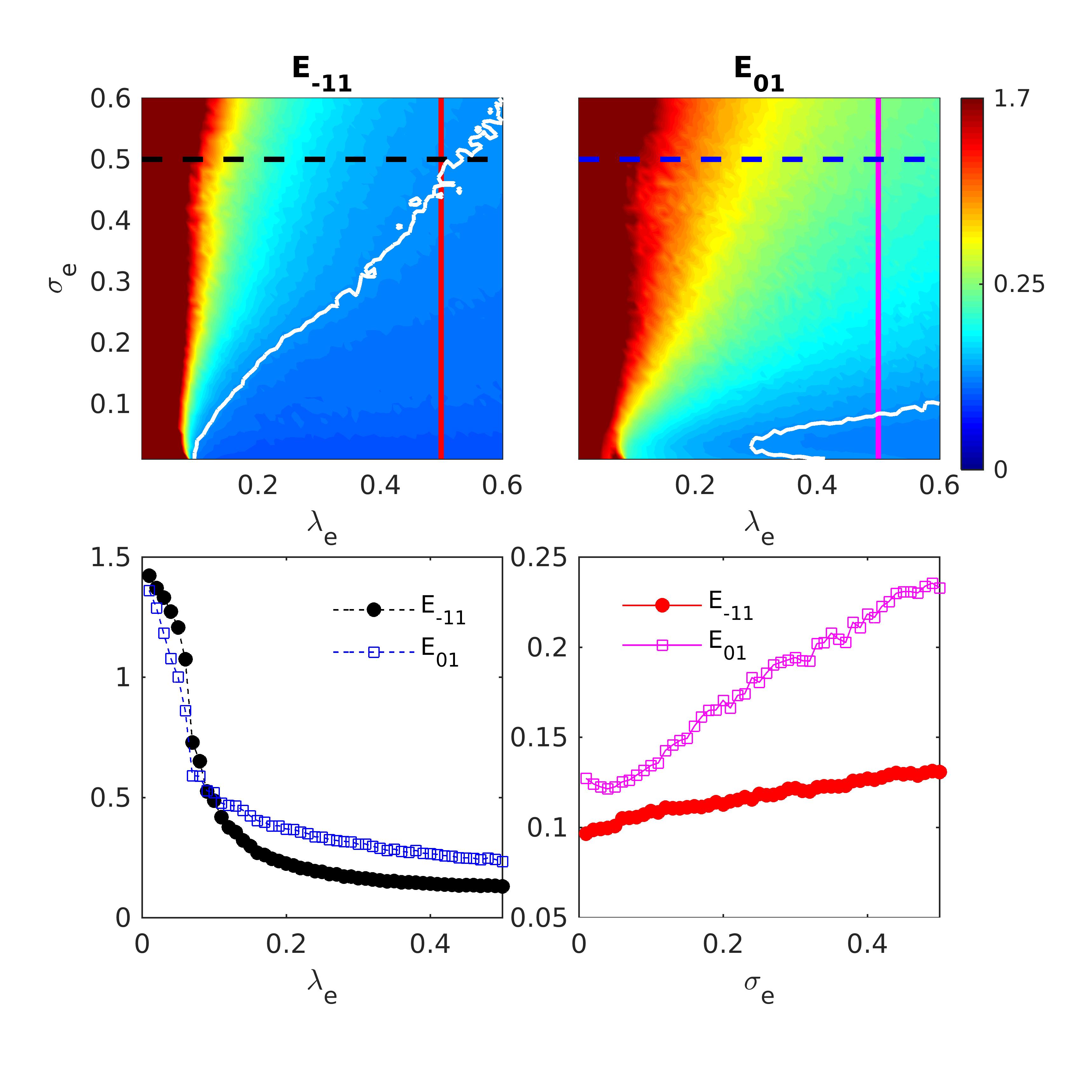}
\caption{Experimental results of relay synchronization in a triplex network with non-identical layers, as a function of the intra-layer ($\sigma_e$) and inter-layer  ($\lambda_e$) couplings. (Top) Colormap of the inter-layer synchronization errors between the outer layers $E_{-11}$ (left) and between one outer layer and the relay layer $E_{01}$ (right) in the $\sigma_e$-$\lambda_e$ parameter space. The white contour line in each panel indicates the isoline for $E_{-11}$ and $E_{01}$ respectively equal to $0.12$, error value taken as a reference. (Bottom) Inter-layer $E_{-11}$, $E_{01}$ synchronization errors as a function of (left) $\lambda_e$ for fixed $\sigma_e=0.5$ (vertical continuous lines in the above panels) and (right) $\sigma_e$ for fixed $\lambda_e=0.5$ (horizontal dashed lines).}
\label{fig:experiment}
\end{figure}

The  experimental results are summarized in Fig.~\ref{fig:experiment}. The top panels represent the averaged experimental inter-layer synchronization error for the outer layers $E_{-11}$ (left)
and between the relay and one of the outer layers $E_{01}$ (right), for all the experimental range of intra-layer $\sigma_e=[0,0.6]$ and inter-layers $\lambda_e=[0,0.6]$ couplings. Even though the system is unavoidably affected by noise and parameter mismatch in the electronic components, for high enough $\lambda_e$ the value of $E_{-11}$ is well below $E_{01}$ and therefore the inter-layer relay synchronization is verified in our experimental setup. Superimposed to the colormaps, we also have drawn the isoline for $E=0.12$ in both panels (white lines), showing that the threshold $\lambda_e^*$ value for which $E_{-11}$ and $E_{01}$ are below the value of the isoline is always smaller in the $E_{-11}$ case.

For a clearer view, in the bottom left panel we have just plotted $E_{-11}$ and $E_{01}$ as a function of $\lambda_e$ for a fixed intra-layer coupling $\sigma_e=0.5$ (corresponding to the blue and black dashed lines in the respective colormap panels in the upper part of Fig.~\ref{fig:experiment}), showing that $E_{-11}$ monotonically goes to zero and is always below $E_{01}$.

Finally, in the bottom-right panel of Fig.~\ref{fig:experiment} we plot both errors, $E_{-11}$ and $E_{01}$, as a function of $\sigma_e$ for a fixed value of the intra-layer coupling $\lambda_e=0.5$ (vertical cuts in red and magenta in the colormap plots). That is done in order to show the effect of increasing the interaction  in the intra-layer connectivity. Similarly to what observed in Fig.~\ref{fig4}, promoting the topological difference between layers as $\sigma_e$ increases rises the synchronization threshold.


\section{Discussion}

Long distance coherence between complex mirrored structures mediated through non-synchronous differentiated ones plays a key role in the functioning of several real-world systems, as for instance the brain. Zero-lag synchronization
has been indeed observed between distant areas of the cortex \cite{Konigqt1997,Soteropoulos2006}, and the transcendental role of symmetry in its dynamics has been lately pointed out \cite{Nicosia2013,Zhang2017}.

In our work we have extended the concept of relay synchronization to the case of a multiplex network, showing that the intermediation of a relay layer can lead to inter-layer synchronization of a set of paired layers, both topologically and dynamically different from the transmitter. The phenomenon  can be extended to indefinitely higher order relay configurations, provided a mirror symmetry is preserved in the multiplex. The coherent state is very robust to changes in the dynamics, topology, and even to strong multiplex disconnection. In this latter scenario, we proved that the low degree nodes in the synchronized outer layers are
responsible for resiilence of the synchronous state, while hubs can be safely de-mutiplexed. Finally, we experimentally validated our results in a multiplex network of three layers of electronic oscillators. 
Our results provide a new path for starting the study of the role of symmetries in setting long distance coherence in real systems.


\section{Acknowledgements}
Work partly supported by the Spanish Ministry of Economy under project FIS2013-41057-P and by GARECOM, Group of Research Excelence URJC-Banco de Santander.  Authors acknowledge the computational resources and assistance provided by CRESCO, the supercomputing center of ENEA in Portici, Italy. R.S.E. acknowledges support from Secretaría de Educación Pública, PRODEP, grant number UDG-PTC-1289-DSA/103.5/16/10313.

\end{document}